\documentclass[letterpaper,11pt]{article}

\usepackage[utf8]{inputenc}

\usepackage[top=3.5cm,bottom=3.5cm, left=4cm,right=4cm,   heightrounded,marginparwidth=1.5cm, marginparsep=1cm]{geometry} 

\usepackage[shortlabels]{enumitem} 

\usepackage[square,numbers,merge,comma,sort&compress]{natbib} 
\makeatletter
\def\NAT@spacechar{\,}  
\makeatother

\usepackage{amsmath,amssymb,amsfonts,amsthm,amsbsy}
\usepackage{mathtools} 

\usepackage{fnpct}
\setfnpct{dont-mess-around} 

\usepackage{slashed,cancel,latexsym}
\usepackage{old-arrows,comment,relsize,setspace,moresize}
\usepackage{epsfig}
\usepackage{changepage,adjustbox}
\usepackage{mathrsfs,calligra,aurical,bold-extra} 
\usepackage{calc,float,appendix}
\usepackage{bigints,xargs,extarrows}
\usepackage{empheq,soul} 

\usepackage[svgnames]{xcolor}  
\definecolor{Blue}{rgb}{0.0, 0.0, 0.37}
\definecolor{Green}{rgb}{0.05, 0.45, 0.25}
\xdefinecolor{dogwoodrose}{rgb}{0.8, 0.1, 0.55}
\xdefinecolor{RRed}{rgb}{0.7, 0.1, 0.525}

\usepackage{tikz}
\usetikzlibrary{scopes,decorations.pathmorphing,patterns,calc,arrows,shapes.geometric,shapes.arrows,decorations.markings,plotmarks}
\usepackage{pgfplots}

\usepackage[blocks]{authblk}  

\usepackage{graphicx} 
\graphicspath{{Figures/}} 

\usepackage[labelsep=colon]{caption}  
\usepackage[labelsep=colon,aboveskip=10pt,belowskip=10pt]{subcaption}
\DeclareCaptionSubType * [roman]{table}
\usepackage{tabularx,array,multirow,makecell,booktabs}  
\usepackage{cellspace}
\newcolumntype{C}[1]{>{\centering\arraybackslash$}S{p{#1}}<{$}}
\newcolumntype{M}[2]{>{\centering\arraybackslash$}#1{#2\linewidth}<{$}}
\newcolumntype{T}[2]{>{\centering\arraybackslash}#1{#2\linewidth}<{}}
\newcolumntype{R}[1]{>{\raggedleft\arraybackslash}m{#1\linewidth}<{}}
\newcolumntype{L}[1]{>{\raggedright\arraybackslash}m{#1\linewidth}<{}}

\newcommand\thinrule{\midrule[0.00001pt]}

\usepackage{titlesec}
\titleformat{\section}{\normalfont\fontsize{12.5}{12}\bfseries}{\thesection}{0.5em}{}
\titleformat{\subsection}{\normalfont\fontsize{10.5}{10}\bfseries}{\thesubsection}{0.5em}{}
\titleformat{\subsubsection}{\normalfont\normalsize\bfseries}{\thesubsubsection}{0.5em}{}

\titlespacing*{\section}{0pt}%
                {4ex plus 1ex minus .5ex}{1.75ex plus .25ex minus .25ex}
\titlespacing*{\subsection}{0pt}%
                {3.5ex plus 1ex minus .5ex}{1.25ex plus .2ex minus .2ex}
\titlespacing*{\subsubsection}{0pt}%
                {2.5ex plus 0.75ex minus .2ex}{0.75ex plus .15ex minus .15ex}
\titlespacing*{\paragraph}{0pt}%
                {1.85ex plus 0.5ex minus .15ex}{1em}

\usepackage{titletoc}
\addtocontents{toc}{\addvspace{-0.75em}}  
\usepackage{bm}  
\usepackage{dsfont}  

\DeclareMathAlphabet{\mathpzc}{OT1}{pzc}{m}{it}
\DeclareMathAlphabet{\mathcal}{OMS}{cmsy}{m}{n}
\DeclareSymbolFontAlphabet{\Scr}{rsfs}
\DeclareMathAlphabet{\mathbold}{U}{BOONDOX-ds}{m}{n}
\SetMathAlphabet{\mathbold}{bold}{U}{BOONDOX-ds}{b}{n}
\DeclareMathAlphabet{\mathcalboondox}{U}{BOONDOX-calo}{m}{n}
\SetMathAlphabet{\mathcalboondox}{bold}{U}{BOONDOX-calo}{b}{n}
\DeclareMathAlphabet{\mathbcalboondox}{U}{BOONDOX-calo}{b}{n}

\DeclareFontFamily{U}{matha}{\hyphenchar\font45}
\DeclareFontShape{U}{matha}{m}{n}{ <5> <6> <7> <8> <9> <10> gen * matha
                    <10.95> matha10 <12> <14.4> <17.28> <20.74> <24.88> matha12}{}
\DeclareSymbolFont{matha}{U}{matha}{m}{n}
\DeclareMathSymbol{\varleftarrow}{3}{matha}{"D0}
\DeclareMathSymbol{\varrightarrow}{3}{matha}{"D1}
\DeclareMathSymbol{\simeq}{3}{matha}{"14}
\DeclareMathSymbol{\sim}{3}{matha}{"12}
\DeclareMathSymbol{\ll}{3}{matha}{"21}
\DeclareMathSymbol{\gtrsim}{3}{matha}{"C1}
\DeclareMathSymbol{\lesssim}{3}{matha}{"C0}

\newcommand\linkcol{RRed}

\usepackage[breaklinks=true,backref=page]{hyperref}
\hypersetup{
    bookmarks=true,         
    bookmarksnumbered=true,
    pdftoolbar=true,        
    pdfmenubar=true,        
    pdffitwindow=false,     
    pdfauthor={},     
    pdfsubject={},   
    pdfcreator={},   
    pdfproducer={},  
    pdfpagemode={UseNone},
    pdfstartview={FitH},
    colorlinks=true,
    plainpages,
    linktoc=page,
    citecolor=blue,
    filecolor=black,
    linkcolor=\linkcol,
    urlcolor=Grey,
}

\newif\ifbackrefshowonlyfirst
\backrefshowonlyfirsttrue
\makeatletter
\let\BR@direct@old@hyper@natlinkstart\hyper@natlinkstart
\renewcommand*{\hyper@natlinkstart}{\phantomsection\BR@direct@old@hyper@natlinkstart}
\let\BR@direct@oldBR@citex\BR@citex
\renewcommand*{\BR@citex}{\phantomsection\BR@direct@oldBR@citex}%

\long\def\hyper@page@BR@direct@ref#1#2#3{\hyperlink{#3}{#1}}

\ifx\backrefxxx\hyper@page@backref
    \let\backrefxxx\hyper@page@BR@direct@ref
    \ifbackrefshowonlyfirst
    \fi
\else
    \ifbackrefshowonlyfirst
    \fi
\fi

\RequirePackage{etoolbox}
\patchcmd{\Hy@backout}{Doc-Start}{\@currentHref}{}{\errmessage{I can't seem to patch backref}}
\makeatother

\renewcommand*{\backref}[1]{}
\renewcommand*{\backrefalt}[4]{%
\ifcase #1 %
\relax
\or
~{\small [\textsc{p.~\fns{\!#2}}]}
\else
~{\small [\textsc{p.~\fns{\!#2}}]}%
\fi}

\usepackage{footnotebackref}
\usepackage{hypernat} 

\usepackage{cleveref} 

\makeatletter
\normalsize
\makeatother

\makeatletter
\g@addto@macro\bfseries{\boldmath}
\makeatother

\protected\def\verythinspace{%
  \ifmmode
    \mskip0.5\thinmuskip
  \else
    \ifhmode
      \kern0.08334em
    \fi
  \fi
}

\newcommand\fns{\footnotesize}

\newcommandx\Hodge[1][1=4,usedefault]{{}^{\star_{#1}}}

\newcommand\Tc{T_\textrm{c}}
\newcommand\ux{\vec{u}_x}
\newcommand\uy{\vec{u}_y}
\newcommand\uz{\vec{u}_z}
\newcommand\E{\mathbf{E}}
\newcommand\B{\mathbf{B}}
\newcommand\Bc{B_\textsc{c}}

\newcommand\Bctwo{B_{\textsc{c}{}_{2}}}
\newcommand\A{\mathbf{A}}

\newcommand\Eg{\mathbf{E}_\textrm{g}}
\newcommand\Bg{\mathbf{B}_\textrm{g}}
\newcommand\Ag{\mathbf{A}_
{\textrm{g}}}

\newcommand\Ezg{\mathbf{E}_{0_{ \textrm{\scriptsize g}}}}
\newcommand\jj{\mathbf{j}}

\newcommand\gstar{g_\star}
\newcommand\Tstar{T^\star}
\newcommand\mstar{m_\star}

\newcommand\phizg{\phi_{0_{ \textrm{\scriptsize g}}}}

\newcommand\mt{\mathrm{m}}

\newcommand\Kelv{\mathrm{K}}

\newcommand{\ms}{\mathsmaller}

\newcommand{\dd}{\partial}
\newcommandx{\mlt}[1]{\mathlarger{\text{#1}}}

\providecommand{\abs}[1]{\left\lvert#1\right\rvert}

\newcommandx{\overbar}[1]{\mkern
1.5mu\overline{\mkern-2.0mu#1\mkern-2.0mu}\mkern 1.5mu}
\newcommandx{\overbarcal}[1]{\mkern6.0mu\overline{\mkern-5.5mu#1\mkern-1.0mu}\mkern 1.5mu}

\title{%
\vspace{-1em}
\centering\boldmath\Large\bfseries%
Gravitational effects in a superconducting film struck by a laser pulse%
\bigskip\bigskip}

\author[1,2]{G.A.\ Ummarino\footnote{\href{mailto:giovanni.ummarino@polito.it}{giovanni.ummarino@polito.it}}}

\author[1]{A.\ Gallerati\footnote{\href{mailto:antonio.gallerati@polito.it}{antonio.gallerati@polito.it}}%
\smallskip}

\affil[1]%
{
{Politecnico di Torino, Dipartimento di Scienza Applicata e Tecnologia, corso Duca degli Abruzzi 24, 10129 Torino, Italy.}%
\medskip\smallskip
}

\affil[2]%
{
{National Research Nuclear University MEPhI, Department of Semiconductor Quantum Electronics, Inst.\ for Physics and Engineering in Biomedicine, Kashira Hwy 31, 115409 Moscow.}%
}

\date{}
\begin{document}
\maketitle
\vspace{-1em}
\begin{abstract}
\noindent
We study the local interaction of the gravitational field with a superfluid condensate. To this end, we exploit the Ginzburg-Landau formalism with generalized Maxwell fields. The analysis shows that a slight local alteration of the gravitational field in a thin superconducting film can be achieved by laser pulses with particular characteristics.
\end{abstract}

\bigskip
\bigskip
\bigskip
\noindent
\begin{tabularx}{\textwidth}{@{}r @{}X}
\textbf{Keywords: } & local affection of gravitational field, superconductivity, gravito-Maxwell, time-dependent Ginzburg-Landau theory.
\end{tabularx}
\bigskip

\section{Introduction} \label{sec:Intro}
In the classical theory of gravitation, as well as in general relativity, the local gravitational field cannot be affected by any medium or device of reasonable density and size. The situation changes, as shown theoretically by Modanese \cite{Modanese:1995tx,Modanese:1996zm}, if the stress-energy tensor is subjected to an appropriate contribution from a suitable macroscopic quantum system (a superconductor or a more general superfluid condensate).\par
In previous works, we have already studied the possibility that supercondensate systems can locally affect the gravitational field
\cite{Ummarino:2017bvz,Ummarino:2019cvw,Ummarino:2020loo,Ummarino:2021vwc,Ummarino:2021tpz,Gallerati:2022nwm} . From an experimental point of view, observable phenomena require a sufficiently long time scale and intensity of the interplay. Our research is then aimed at studying particularly favorable situations, where the intensity of the effect and its duration are maximized. In this regard, we are going to explore the effects of particular electromagnetic fields on a supercondensate immersed in a weak gravitational field. With respect to our previous analyses, we will exploit the additional effects of incident laser beams, bringing the system into a more convenient condition for experimental observations.
\par
As one might expect, the increase in system complexity is reflected in troubles with the mathematical formalism describing the interplay.
To study the new setting with a tractable mathematical formulation, we will consider a simple system consisting of a thin superconducting film, with thickness smaller than the superconductive penetration depth. The latter film is then hit, orthogonally to its center, by a laser beam with suitable time-dependent frequence and intensity. This will generate an external vector potential and associated generalized electric field, driving the effective interaction between the supercondensate and the local gravitational field.
The chosen setup allows to neglect the spatial dependence of the physical quantities involved, so that the behavior of the system can be described in terms of time-dependent differential equations of the first order. This also provides a simpler framework in which to determine the most favorable situation for a localized interaction with the gravitational field.\par
The paper is organized as follows. In Section \ref{sec:gravMax} we briefly describe the gravito-Maxwell formalism, giving rise to the generalized Maxwell fields. In Section \ref{sec:model} we give an explicit formulation for the physical interaction by means of a generalized time-dependent Ginzburg Landau (TDGL) theory, taking into account the effects of the laser pulses.
In Section \ref{sec:discuss} we analyse the experimental predictions about the local gravitational affection and study how to determine the most favorable situation for a macroscopic effect; this, in turn, requires an optimization of the input parameters in the Ginzburg-Landau equations, in order to maximize the interaction. Finally, in Section \ref{sec:concl} we summarize our results and give some insights about possible future developments.

%

\section{Gravito-Maxwell formalism}
\label{sec:gravMax}
Our starting point for a well-defined model, leading to explicit experimental predictions, is the appearance of generalized fields and potentials in superconductors, induced by the presence of a local gravitational field coupled to the supercondensate \cite{papini1969gravity,li1991effects,Anandan:1983fe,torr1993gravitoelectric,agop2000local,Cabral:2016klm,Tajmar:2004ww,Tajmar:2006gh,tajmar2008electrodynamics,Ruggiero:2021uag,Toth:2021dut,Inan:2017qdt,inan2022superconductor}. The latter phenomenon can be suitably characterized by an effective model, where the precise definition of the generalized fields emerge from a weak-field expansion for the Earth local gravitational field. In the following, we briefly summarize the mathematical formulation of this gravito-Maxwell approach \cite{Ummarino:2017bvz,Gallerati:2022pgh}.\par
Let us consider a weak gravitational background, where the (nearly-flat) spacetime metric $g_{\mu\nu}$ is expressed as
\begin{equation}
g_{\mu\nu}\:\simeq\:\eta_{\mu\nu}+h_{\mu\nu}\:,
\end{equation}
with $\eta_{\mu\nu}=\mathrm{diag}(-1,+1,+1,+1)$ and where $h_{\mu\nu}$ is a small perturbation of the flat Minkowski spacetime.
\footnote{%
Here we work in the `mostly plus' convention and natural units $c=\hbar=1$\,.
}
If we now introduce the symmetric traceless tensor
\begin{equation}
\bar{h}_{\mu\nu}=h_{\mu\nu}-\frac12\,\eta_{\mu\nu}\,h\:,
\end{equation}
with $h=h^\sigma{}_\sigma$, it can be shown that the Einstein equations in the harmonic De Donder gauge $\partial^{\mu}\bar{h}_{\mu\nu}\simeq0$ can be rewritten, in linear approximation, as \cite{Ummarino:2017bvz,Ummarino:2019cvw,Ummarino:2020loo,Gallerati:2020tyq,Ummarino:2021vwc,Gallerati:2021ops}
\begin{equation}
R_{\mu\nu}-\frac12\,g_{\mu\nu}\,R=\partial^{\rho}\mathscr{G}_{\mu\nu\rho}\:=\:8\pi\mathrm{G}\,T_{\mu\nu}\:,
\end{equation}
having also defined the tensor
\begin{equation}
\mathscr{G}_{\mu\nu\rho}\:\equiv\:\partial_{{[}\nu}\bar{h}_{\rho{]}\mu}+\partial^{\sigma}\eta_{\mu{[}\rho}\,\bar{h}_{\nu{]}\sigma}
     \:\simeq\:\partial_{{[}\nu}\bar{h}_{\rho{]}\mu}\:.
\end{equation}
%
We then introduce the fields \cite{agop2000local,Ummarino:2017bvz}
\begin{align}
\Eg=-\frac12\,\mathscr{G}_{00i}=-\frac12\,\partial_{{[}0}\bar{h}_{i{]}0}\,,
\qquad
\Ag=\frac14\,\bar{h}_{0i}\,,
\qquad
\Bg=\frac14\,{\varepsilon_i}^{jk}\,\mathscr{G}_{0jk}\,,
\end{align}
for which we get, restoring physical units, the set of equations \cite{agop2000local,Ummarino:2017bvz,Ummarino:2019cvw,Behera:2017voq}
\begin{equation}
\label{eq:gravitoMaxwell}
\begin{alignedat}{2}
&\nabla\cdot\Eg=-4\pi\mathrm{G}\,\rho_\text{g}\,,\qquad\; &
&\nabla\cdot\Bg=0 \,,
\\[2.5\jot]
&\nabla\times\mathbf{E}_\text{g}=-\dfrac{\partial\mathbf{B}_\text{g}}{\partial t}\,,\qquad\; &
&\nabla\times\mathbf{B}_\text{g}=-4\pi\mathrm{G}\,\frac{1}{c^2}\,\mathbf{j}_\text{g}
    +\frac{1}{c^2}\,\frac{\partial\mathbf{E}_\text{g}}{\partial t}\,,
\end{alignedat}
\end{equation}
\sloppy
having defined, in a comoving reference frame, the mass density \,$\rho_\text{g}\equiv T_{00}$\, and the mass current density \,${\mathbf{j}_\text{g}\equiv- T_{0i}}$\,.%
\footnote{%
The latter identifications differ by a sign with respect to our previous works, and determine minus signs in \eqref{eq:gravitoMaxwell} and subsequent \eqref{eq:generalizedMaxwell}.}
The above equations have the same formal structure of the Maxwell equations, with $\mathbf{E}_\textrm{g}$ and $\mathbf{B}_\textrm{g}$ gravitoelectric and gravitomagnetic field, respectively.
\par

\paragraph{Generalized fields and equations.}
Now we introduce generalized electric/magnetic fields, scalar and vector potentials, featuring both electromagnetic and gravitational contributions \cite{agop2000local,agop2000some,Ummarino:2017bvz,Ummarino:2019cvw}:
\begin{equation}
\mathbf{E}=\mathbf{E}_\text{e}+\frac{m}{e}\,\mathbf{E}_\text{g}\,,\qquad
\mathbf{B}=\mathbf{B}_\text{e}+\frac{m}{e}\,\mathbf{B}_\text{g}\,,\qquad
\phi=\phi_\text{e}+\frac{m}{e}\,\phi_\text{g}\,,\qquad
\mathbf{A}=\mathbf{A}_\text{e}+\frac{m}{e}\,\mathbf{A}_\text{g}\,,
\label{eq:genfields}
\end{equation}
where $m$ and $e$ identify the mass and electron charge, respectively.
The generalized Maxwell equations for the new fields read \cite{agop2000local,Ummarino:2017bvz,Ummarino:2019cvw,Ummarino:2020loo,Gallerati:2021ops,Ummarino:2021vwc}:
\begin{equation}
\label{eq:generalizedMaxwell}
\begin{alignedat}{2}
&\nabla\cdot\mathbf{E}=\left(\frac{1}{\varepsilon_0}-\frac{1}{\varepsilon_\text{g}}\right)\rho\,,\qquad\;&
&\nabla\cdot\mathbf{B}=0 \,,
\\[2.5\jot]
&\nabla\times\mathbf{E}=-\dfrac{\partial\mathbf{B}}{\partial t}\,,\qquad\;&
&\nabla\times\mathbf{B}=(\mu_0-\mu_\text{g})\,\mathbf{j}
    +\frac{1}{c^2}\,\frac{\partial\mathbf{E}}{\partial t}\,,
\end{alignedat}
\end{equation}
where $\varepsilon_0$ and $\mu_0$ are the vacuum electric permittivity and magnetic permeability. In the above equations, $\rho$ and $\mathbf{j}$ identify the electric charge density and electric current density, respectively, while the mass density and the mass current density vector can be expressed in terms of the latter as
\begin{equation}
\rho_\text{g}=\frac{m}{e}\,\rho\,,\qquad\quad
\mathbf{j}_\text{g}=\frac{m}{e}\:\mathbf{j}\:,
\end{equation}
while the vacuum \emph{gravitational} permittivity $\varepsilon_\text{g}$ and permeability $\mu_\text{g}$ read
\begin{equation}
\varepsilon_\text{g}=\frac{1}{4\pi\mathrm{G}}\,\frac{e^2}{m^2}\:,\qquad\quad
\mu_\text{g}=\frac{4\pi\mathrm{G}}{c^2}\,\frac{m^2}{e^2}\:.
\end{equation}

\smallskip


\section{The model}
\label{sec:model}

\subsection{Time-dependent Ginzburg-Landau formulation}
Let us consider a superconducting sample on the Earth surface, in the presence of an external magnetic field $\B_0$. The standard Ginzburg-Landau equations describing the system can be written as \cite{tang1995time,fleckinger1998dynamics,kopnin1999time}:
\begin{subequations}
\label{eq:TDGL0}
\begin{align}
&\frac{\hbar^2}{2\,\mstar}\left(i\,\nabla
        +\frac{2\,e}{\hbar}\,\A\right)^{\!2}\psi -a\,\psi+b\,\abs{\psi}^2\psi
        \;=\,-\frac{\hbar^2}{2\,\mstar\,\mathcal{D}}\left(\frac{\dd}{\dd t}
                +\frac{2\,i\,e}{\hbar}\,\phi\right)\,\psi\;,
\\[2\jot]
&\nabla\times\nabla\times\A-\nabla\times\B_0=\mu_0\,\big(\jj_\text{n}+\jj_\text{s}\big)\,,
\end{align}
\end{subequations}
where $\mstar$ is the mass of a Cooper pair, $\mathcal{D}$ the diffusion coefficient, $\sigma$ the conductivity in the normal phase.
We already pointed out that the interaction with the local weak gravitational field leads to the appearance of effective generalized Maxwell fields \eqref{eq:genfields}, so the above equation refer to generalized external magnetic field $\B_0$ and vector potential $\A$, featuring both electromagnetic and gravitational contributions.
The $a$ and $b$ coefficients read
\begin{equation}
a \equiv a(T)=a_{0}\,(T-\Tc)\,,\qquad\qquad
b \equiv b(\Tc)\,,\qquad
\end{equation}
$a_0$, $b$ being positive constants.  The contributions related to the normal current and supercurrent densities can be explicitly written as
\begin{equation}
\label{eq:currents}
\begin{split}
\jj_\text{n}&=-\sigma\left(\frac{\dd\A}{\dd t}+\nabla\phi\right)\:,
\\[3\jot]
\jj_\text{s}&=-i\,\hbar\,\frac{e}{\mstar}\left(\psi^*\,\nabla\psi-\psi\,\nabla\psi^*\right)
    -\frac{4\,e^2}{\mstar}\,\abs{\psi}^2\A\:.
\end{split}
\end{equation}
The boundary and initial conditions are
\begingroup%
\setlength{\belowdisplayskip}{1pt plus 1pt minus 4pt}%
\setlength{\belowdisplayshortskip}{1pt plus 1pt minus 4pt}%
\begin{align}
  \left.
  \begin{aligned}
  \left(i\,\nabla\psi+\frac{2\,e}{\hbar}\,\A\,\psi\right)\cdot\mathbf{n}=0&  \cr
  \hfill\nabla\times\A\cdot\mathbf{n}=\B_0\cdot\mathbf{n}&  \\[1.5\jot]
  \hfill\A\cdot\mathbf{n}=0&
  \end{aligned}
  \;\;\right\} \; \text{on }\dd\Omega\times(0,t)\;,
\qquad\quad
  \left.
  \begin{aligned}
  \psi(x,0)&=\psi_0(x) \cr
  \A(x,0)&=\A_0(x)
  \end{aligned}
  \!\!\!\!\right\} \; \text{on }\Omega\;,\qquad
\label{eq:boundary}
\end{align}
\endgroup
where $\dd\Omega$ is the boundary of a smooth and simply connected domain in $\mathbb{R}^\ms{\textrm{N}}$. We denote by $\A_0$ and $\phi_0$ the external vector and scalar potentials, coinciding with the internal values when the sample is in the normal state and the material is very weakly diamagnetic. Finally, $\psi_0(x)$ is the order parameter in the unperturbed superfluid state.

\paragraph{Dimensionless Ginzburg-Landau equations.}
Let us put ourselves in the Coulomb gauge \,$\nabla\cdot\A=0$\, so that, when $\B_0$ is uniform,
\begin{equation}
\nabla^2\A=-\mu_0\left(\jj_\text{n}+\jj_\text{s}\right)\:.
\end{equation}
In order to write eqs.\ \eqref{eq:TDGL0} in a dimensionless form, we introduce the quantities
\begingroup%
\begin{equation} \label{eq:param}
\begin{split}
&\psi_0^2(T)=\frac{\abs{a(T)}}{b}\,,
\quad\;
\xi(T)=\frac{\hbar}{\sqrt{2\,\mstar\abs{a(T)}\,}}\,,
\quad\;
\lambda(T)=\sqrt{\frac{b\,\mstar}{4\,\mu_0\abs{a(T)} e^2\,}}\,,
\quad\;
k=\frac{\lambda(T)}{\xi(T)}\,,
\\[2\jot]
&\tau(T)=\frac{\lambda^2(T)}{\mathcal{D}}\,,
\quad\quad
\eta=\mu_0\,\sigma\,\mathcal{D}\,,
\quad\quad
\Bc(T)=\sqrt{\frac{\mu_0\abs{a(T)}^2}{b}}=
    \frac{\hbar}{2\sqrt{2}\,e\:\lambda(T)\,\xi(T)}\,,
\end{split}
\end{equation}
\endgroup
where $\lambda(T)$, $\xi(T)$ and $\Bc(T)$ are the penetration depth, coherence length and thermodynamic critical field, respectively. We also define the dimensionless quantities
\begin{equation}
t'=\:\frac{t}{\tau}\,,\qquad
x'=\:\frac{x}{\lambda}\,,\qquad\;
y'=\:\frac{y}{\lambda}\,,\qquad\;
z'=\:\frac{z}{\lambda}\,,\qquad\;
\psi'=\:\frac{\psi}{\psi_0}\,,\quad
\label{eq:dimlesscoord}
\end{equation}
and the dimensionless potentials, fields and currents can be then expressed as:
\begingroup%
\begin{equation}
\A'=\frac{\A\,k}{\sqrt{2}\,\Bc\,\lambda}\,,
\quad\:
\phi'=\frac{\phi\,k}{\sqrt{2}\,\Bc\,\mathcal{D}}\,,
\quad\:
\E'=\frac{\E\,\lambda\,k}{\sqrt{2}\,\Bc\,\mathcal{D}}\,,\quad
\B'=\frac{\B\,k}{\sqrt{2}\,\Bc}\,,
\quad\:
\jj'=\frac{\jj\,\mu_0\,\lambda\,k}{\sqrt{2}\,\Bc}\,.
\label{eq:dimlessfields}
\end{equation}
\endgroup
Finally, we define the dimensionless parameter \cite{Ummarino:2017bvz}
\begin{equation}
\gstar=\frac{\lambda(T)\,k\,\mstar\,g}{\sqrt{2}\,e\,\Bctwo\,\mathcal{D}} \ll 1\:,
\end{equation}
proportional to the Earth surface gravity acceleration $g$.
\par\smallskip
Using \eqref{eq:dimlesscoord} and \eqref{eq:dimlessfields} in eqs.\ \eqref{eq:currents} and \eqref{eq:TDGL0} we find the dimensionless time-dependent Ginzburg-Landau (TDGL) equations in a bounded, smooth and simply connected domain in $\mathbb{R}^\ms{\textrm{N}}$ \cite{tang1995time,lin1997ginzburg}%
\footnote{%
From now on, we drop the primes for the sake of notational simplicity.%
}:
\begin{subequations}
\label{eq:dimlessTDGL}
\begingroup%
\setlength{\belowdisplayskip}{0pt plus 3pt minus 4pt}%
\setlength{\belowdisplayshortskip}{0pt plus 3pt minus 4pt}%
\begin{align}
&\frac{\dd\psi}{\dd t} \,+\, i\,\phi\,\psi
    \,+\,k^2\left(\abs{\psi}^2-1\right)\,\psi
    \,+\,\left(i\,\nabla+\A\right)^2 \psi=0 \;,
\label{subeq:dimlessTDGLn1}
\\[2\jot]
&\nabla\times\nabla\times\A\,-\,\nabla\times\B_0
    =\jj_\text{n}+\jj_\text{s}
    \:=\,-\,\eta\,\left(\frac{\dd\A}{\dd t}+\nabla\phi\right)
        -\frac{i}{2}\left(\psi^*\nabla\psi-\psi\nabla\psi^*\right)
        -\abs{\psi}^2\A\,,
\label{subeq:dimlessTDGLn2}
\end{align}
\endgroup
\end{subequations}
$k$, $\eta$ e $\A_{0}$ being the three parameters describing our system. The boundary and initial conditions \eqref{eq:boundary} become, in the dimensionless form,
\begin{align}
  \left.
  \begin{aligned}
  \left(i\,\nabla\psi+\A\,\psi\right)\cdot\mathbf{n}=0&\cr
  \nabla\times\A\cdot\mathbf{n}=\B_0\cdot\mathbf{n}&\cr
  \A\cdot\mathbf{n}=0&
  \end{aligned}
  \!\!\!\right\} \; \text{on }\dd\Omega\times(0,t)\;;
\qquad\;
  \left.
  \begin{aligned}
  \psi(x,0)&=\psi_0(x)\cr
  \A(x,0)&=\A_0(x)
  \end{aligned}
  \!\!\!\!\right\} \; \text{on }\Omega\;.\quad
\label{eq:dimlessboundary}
\end{align}
In the (dimensionless) Coulomb gauge, \,$\nabla\cdot\A=0$, the above \eqref{eq:dimlessTDGL} read
\begin{subequations}
\label{eq:dimlessTDGLCoulgauge}
\begin{align}
&\frac{\partial \psi}{\partial t}+i\,\phi\,\psi+k^{2}\left(\abs{\psi}^{2}-1\right)\,\psi+\big(-\Delta\psi+i\,\textbf{A}\cdot\nabla\psi\big)+\A^{2}\,\psi=0\:,
\label{subeq:dimlessTDGLCoulgauge1}
\\[0.75ex]
-&\Delta\textbf{A}=-\eta\left(\frac{\partial \textbf{A}}{\partial t}+\nabla\phi\right)-\frac{i}{2}\left(\psi^{*}\,\nabla\psi-\psi\,\nabla\psi^{*}\right)-\abs{\psi}^{2}\A\:.
\label{subeq:dimlessTDGLCoulgauge2}
\end{align}
\end{subequations}

\subsection{Laser pulse}
We now examine in detail the experimental setting introduced in Sect.\ \ref{sec:Intro} and the relevant physical implications.\par
\begin{figure}[H]
\captionsetup{skip=1pt,belowskip=5pt,aboveskip=20pt,font=small,labelfont=small,format=hang}
\centering
\includegraphics[keepaspectratio, width=0.525\textwidth]{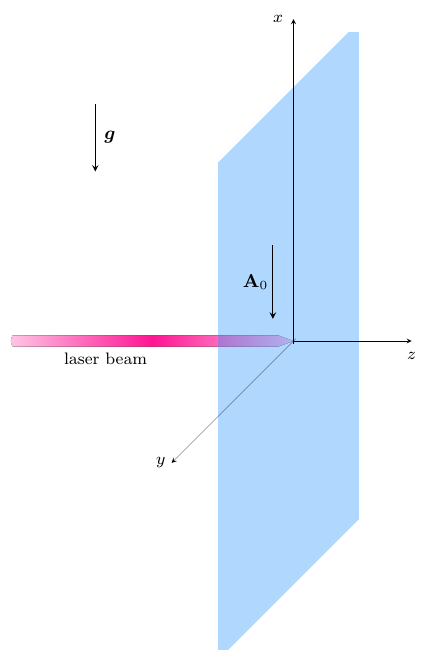}
\caption{Schematic representation of the system under consideration: a superconducting thin film is excited by a normally incident, $x$-polarized electromagnetic pulse.}
\label{fig:laser}
\end{figure}

Let us consider an horizontal thin superconducting film of thickness $L$,
see Figure \ref{fig:laser}. The origin of a cartesian reference system is put at the center of the film, at mid-thickness, with $\uy$, $\uz$ parallel to the ground and $\ux$ along the vertical direction and perpendicular to the film surface. At \,$t=0$, the film is struck by an orthogonal laser pulse, giving rise to an external electromagnetic vector potential $\A_0(t)$ of the form%
\footnote{%
From now on, we denote by a zero subscript the values of the initial external potentials and fields.}
\begin{equation}
\label{eq:A0}
\A_0 =-A_0\:\theta(t)\:\ux\:,
\end{equation}
where $\theta(t)$ is the Heaviside function.\par

\medskip
Then, we consider the presence of the constant, external Earth gravitational field along the vertical direction. The latter constitutes the gravitational component $\Ezg=-\gstar\,\ux$\, of an external generalized electric field.%
\footnote{%
In dimensional units, \,$\Ezg=-\frac{m}{e}\,g\:\ux$\,.
}
We then write
\begingroup
\setlength{\abovedisplayskip}{6pt plus 3pt minus 4pt}%
\setlength{\belowdisplayskip}{8pt plus 3pt minus 4pt}%
\begin{equation}
\label{eq:E0g}
\Ezg={E}_{0_{\scriptstyle\textrm{g}}}\,\ux\:=-\gstar\,\ux=-\nabla \phizg\:,
\end{equation}
\endgroup
with a related scalar potential $\phizg=\gstar\,x$.\par\smallskip
Following the discussion of Sect.\ \ref{sec:gravMax}, the external generalized field $\E_0$ is then expressed as
\begin{equation}
\label{eq:E0}
\E_0=-\nabla\phizg-\frac{\partial \A_0}{\partial t}={E}_{0_{\scriptstyle\textrm{g}}}\,\ux+A_0\,\delta(t)\,\ux\:.
\end{equation}
Let us now consider equations \eqref{eq:dimlessTDGLCoulgauge}. In order to find
an intensity scale at which possible gravitational effects in the superconductor become observable, we make the assumption that the order parameter $\psi$ varies slowly in the spatial variables, being the film thin and the perturbations homogeneous. This in turn allows to neglect the spatial dependence of $\psi$ \cite{Biancalana2017}, allowing for a simplification of \eqref{eq:dimlessTDGLCoulgauge}. Equation \eqref{subeq:dimlessTDGLCoulgauge1} can be then rewritten as
\begin{equation}
\frac{\partial \psi}{\partial t}+i\,\phi\,\psi+k^{2}\left(|\psi|^{2}-1\right)\psi+A^{2}\,\psi=0\:.
\label{eq:newdimlessTDGL1}
\end{equation}
In the absence of the laser pulse and neglecting the gravitational field we would have
\begin{equation}
k^{2}\left(|\psi|^{2}-1\right)=0\:,
\end{equation}
that implies, for these conditions,  $\psi=\psi_{0}=1$.
We then look for general solutions inside the superconductor of the form
\begin{equation}
\begin{split}
\psi(t)&=\psi_{0}+\psi_{1}(t)+\gstar\,\psi_{2}(t)\:,
\\[1.25ex]
A(t)&=A_{1}(t)+\gstar\,A_{2}(t)\:,
\\[1.25ex]
\phi&=\phizg=\gstar\,x\:,
\label{eq:fieldexpans}
\end{split}
\end{equation}
where we stopped the expansion at first order in $\gstar$, being $\gstar\ll1$. Here, $A(t)$ denotes the magnitude of the vector potential modified by the presence of the superconductor. The terms
$\psi_{1}(t)$ and $A_{1}(t)$ are connected to the perturbation of the system due to the presence of the laser, while $\psi_{2}(t)$ and $A_{2}(t)$ are determined by the presence of the gravitational field.
In this regard, we are interested in the laser effect on the superconducting condensate, analysing a possible enhance of the local perturbation on the gravitational field.\par\smallskip
Let us now derive the equations for $\psi_{1}(t)$, $\psi_{2}(t)$, $A_{1}(t)$ and $A_{2}(t)$. To this end, we will analyse the contributions in the TDGL equations at different orders in $\gstar$.\par
Inserting \eqref{eq:fieldexpans} in \eqref{eq:newdimlessTDGL1} we obtain for $\psi_1$ the relation
\begin{equation}
\label{eq:dpsi1}
\frac{\partial\psi_{1}(t)}{\partial t}+k^{2}\,\psi_{1}^{3}(t)+3\,k^{2}\,\psi_{1}^{2}(t)+\left(k^{2}+A_{1}^{2}(t)\right)\psi_{1}(t)=-A_{1}^{2}(t)\:,
\end{equation}
at zero-order in $\gstar$, with the initial condition $\psi_{1}(0)=0$.\par\smallskip
The variation of the $\psi_2$ component comes from the expansion at first-order in $\gstar$ and reads
\begin{equation}
\frac{\partial\psi_{2}(t)}{\partial t}+\left(3\,k^{2}\big(1+\psi_{1}(t)\big)^{2}-k^{2}+A_{1}^{2}(t)\right)\psi_{2}(t)=-\big(2\,A_{1}(t)\,A_{2}(t)+i\,x\big) \big(1+\psi_{1}(t)\big)\,.
\end{equation}
We assume that, when the laser is turned on at $t=0$, the superconductor has already been subjected to the external static gravitational field for a sufficiently long time. This ensures that the effect of any gravitational transient on the supercondensate has already vanished \cite{Ummarino:2017bvz} so that, before the laser pulse, the system is in equilibrium. We are therefore exploring physical effects other than \cite{Ummarino:2017bvz,Ummarino:2021vwc,Ummarino:2021tpz}, since the local alteration does not arise from the transition of the sample to the superconducting state, but from its interaction with the laser pulse at subsequent times.\par
In order to simplify calculations, we study what happens at the center of the superconductor, corresponding to $x=0$. This gives for $\psi_2$
\begin{equation}
\label{eq:dpsi2}
\frac{\partial \psi_{2}(t)}{\partial t}+\left(3\,k^{2}\big(1+\psi_{1}(t)\big)^{2}-k^{2}+A_{1}^{2}(t)\right)\psi_{2}(t)=-2\,A_{1}(t)\,A_{2}(t)\,\big(1+\psi_{1}(t)\big)\:,
\end{equation}
with the initial condition $\psi_{2}(0)=0$.\par\smallskip
Using \eqref{eq:fieldexpans} in \eqref{subeq:dimlessTDGLCoulgauge2}, from the zero-order in $\gstar$ we obtain the variation of the $A_1$ contribution
\begin{equation}
\label{eq:dA1}
\eta\:\frac{\partial A_{1}(t)}{\partial t}=-A_{1}(t)\,\big(1+\psi_{1}(t)\big)^{2}\:,
\end{equation}
with the initial condition  $A_{1}(0)=-A_{0}$, see \eqref{eq:A0}.\par\smallskip
Finally, from the contribution at first-order in $\gstar$, we find for $A_2$
\begin{equation}
\label{eq:dA2}
\eta\left(\frac{\partial A_{2}(t)}{\partial t}+1\right)=-A_{2}(t)\,\big(1+\psi_{1}(t)\big)^{2}-2\,A_{1}(t)\,\psi_{2}(t)\,\big(1+\psi_{1}(t)\big)\:,
\end{equation}
with the initial condition  $A_{2}(0)=-\eta$. \,The latter condition is found starting from eq.\ \eqref{eq:E0} at first order in $\gstar$
\begin{equation}
-\,\frac{\partial\phizg}{\partial x}-\gstar\left.\frac{\partial A_{2}}{\partial t}\right|_{t=0}\!\!\!\:=\:-\gstar
\quad\;\Rightarrow\;\quad
\left.\frac{\partial A_{2}}{\partial t}\right|_{t=0}=0\:,
\end{equation}
so that from \eqref{eq:dA2} we obtain
\begin{equation}
0\:=\:\eta
\left.\frac{\partial A_{2}}{\partial t}\right|_{t=0}\!\!
=\:-A_{2}(0)\,\big(1+\xcancel{\psi_{1}(0})\big)^{2}-2\,A_{1}(0)\,\xcancel{\psi_{2}(0})\,\big(1+\psi_{1}(0)\big)-\eta\:.\quad
\end{equation}\par
\smallskip
To solve the resulting system of four equations, we initially put in equations \eqref{eq:dpsi1} and \eqref{eq:dpsi2} the initial conditions
\begin{equation}
\begin{split}
A_{1}(0)&=-A_{0}\,,\\[\jot] 
A_{2}(0)&=-\eta\,,
\end{split}
\end{equation}
and then we solve for $\psi_{1}(t)$ and $\psi_{2}(t)$.
Subsequently, we use the latter results to obtain $A_{1}(t)$ and $A_{2}(t)$ from \cref{eq:dA1,eq:dA2}.
The process converges already at the first iteration.\par
In the special case under consideration, with a laser impulse of the form \eqref{eq:A0}, it is possible to find cumbersome analytical solutions for $\psi_1(t)$, $\psi_2(t)$, $A_1(t)$, $A_2(t)$, that we report in Appendix \ref{app:sol}.

\section{Discussion}
\label{sec:discuss}
\sloppy
From a general point of view, when solving the Ginzburg-Landau equations describing our framework, it is necessary to respect the physical prescription
\begin{equation}
A_0 < k\:,
\label{eq:A0_k}
\end{equation}
since the condition $A_0 \geq k$\, would cause the system under consideration to return to its normal state.\par
\begin{figure}[H]
\captionsetup{skip=15pt,belowskip=15pt,font=small,labelfont=small,format=hang}
\centering
\begin{subfigure}[t]{\textwidth}
\includegraphics[width=0.95\textwidth,keepaspectratio]{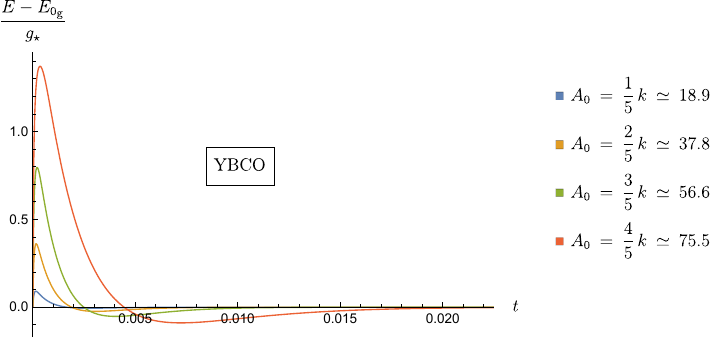}
\label{subfig:YBCO_A0small}
\end{subfigure}
\par\vspace{1.75cm}
\begin{subfigure}[b]{\textwidth}
\includegraphics[width=1.025\textwidth,keepaspectratio]{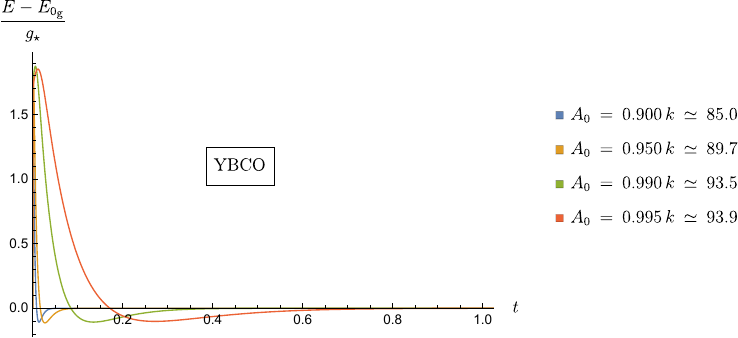}
\label{subfig:YBCO_A0big}
\end{subfigure}
\caption{Gravitational field alteration as a function of time (adimensional units) for an YBCO film.}
\label{fig:YBCO}
\end{figure}
\begin{figure}[H]
\captionsetup{skip=15pt,belowskip=15pt,font=small,labelfont=small,format=hang}
\centering
\begin{subfigure}[t]{\textwidth}
\includegraphics[width=0.95\textwidth,keepaspectratio]{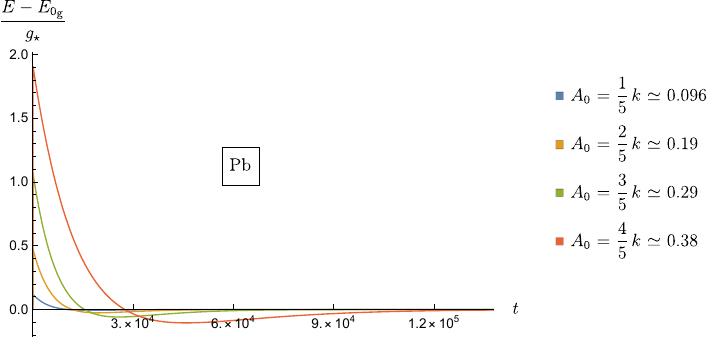}
\label{subfig:Pb_A0small}
\end{subfigure}
\par\vspace{1.75cm}
\begin{subfigure}[b]{\textwidth}
\includegraphics[width=1.02\textwidth,keepaspectratio]{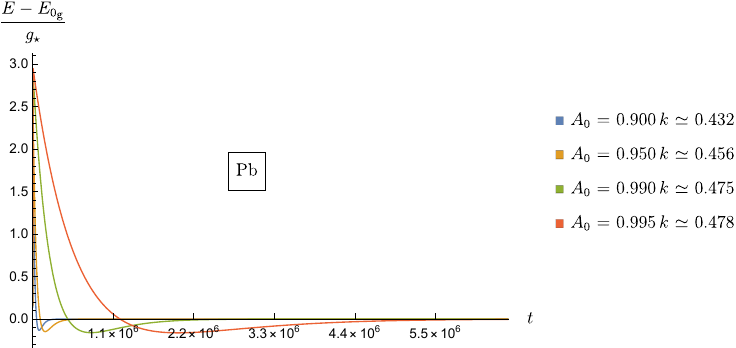}
\label{subfig:Pb_A0big}
\end{subfigure}
\caption{Gravitational field alteration as a function of time (adimensional units) for a Pb film.}
\label{fig:Pb}
\end{figure}

Figures \ref{fig:YBCO} and \ref{fig:Pb} show the calculated local alteration of the gravitational field, for different values of $A_0$, for two superconductors, YBCO and Pb. The former is an high critical temperature superconductor of the second type, while the latter is low critical temperature superconductor of the first type. In particular, we study the magnitude of the ratio
\begin{equation}
\frac{\E-\Ezg}{\gstar}\:,
\end{equation}
where $\Ezg$ is given by the constant contribution of the standard Earth gravitational field \eqref{eq:E0g}, while $\E$ is given
by
\begin{equation}
\label{eq:Efin}
\E=-\nabla\phi-\frac{\partial \A}{\partial t}\:,
\end{equation}
and represents the perturbed generalized field, affected by the presence of the superconductor, whose form is found using the solutions \eqref{eq:fieldexpans} of the Ginzburg-Landau equation.\par
Every variation is calculated at a sample temperature $\Tstar$ such that the ratio
\begin{equation}
\frac{\Tc-\Tstar}{\Tc}
\end{equation}
is approximately the same in the two materials. The latter $\Tstar$ is chosen so that the material is in the superconducting interval of temperatures where the Ginzburg-Landau formulation apply and a mean field approximation can be exploited. The described situation takes place for a range of temperatures close to the critical $\Tc$, but sufficiently far from it so as to prevent the appearance of fluctuations.
The results show that the local gravitational affection is maximized for values of $A_0$ that are close to $k$, and the extent of the response is similar for the two superconductors.\par
Restoring dimensional quantities through \eqref{eq:dimlessfields} we would find
\begin{equation}
A_0 \quad\rightarrow\quad\: A_0\:\frac{k}{\sqrt{2}\,\Bc\,\lambda}=\frac{A_0}{A_n}\,,\qquad
A_n=\frac{\sqrt{2}\,\Bc\,\lambda}{k}\,,\quad
\end{equation}
so that the dimensional version of condition \eqref{eq:A0_k}, involving the maximum allowed value for the vector potential magnitude, is given by
\begin{equation}
\frac{A_0}{A_n}<k
\quad\Rightarrow\quad
A_0<\sqrt{2}\,\Bc\,\lambda=A_0^{{}^{\textsc{max}}}
\end{equation}
then depending on penetration depth and thermodynamic critical field.\par
Apparently, there is a great difference in the characteristic times of the phenomenon for the two superconductors. However,  restoring again dimensional quantities through \cref{eq:dimlesscoord,eq:dimlessfields}, we roughly obtain comparable time intervals, of the order of $10^{-9}\,\mathrm{s}$. The latter result can be directly obtained multiplying by time units $\tau$ (from Table \ref{tab:parameters}) the time scales deduced from the plots.\par
We can also see from Figures \ref{fig:YBCO} and \ref{fig:Pb} that an initial time range exists where the altered gravitational field receives a positive contribution. After this phase, there is a time interval where the field is decreased with respect to the standard external intensity. Finally, the field relaxes returning to the unperturbed value.

\subsection{Oscillating vector potential}
Now we examine the effects originating from the presence of an oscillating laser pulse. We study the particularly simple case
\begin{equation}
\label{eq:A0cos}
\A_0(t)=A_0\,\cos\left(\omega\,t\right)\:\ux\:.
\end{equation}
In this case, it is no longer possible to obtain affordable analytical solutions of the system given by the Ginzburg-Landau equations \eqref{eq:dpsi1}--\eqref{eq:dA2} for the fields \eqref{eq:fieldexpans}, but the behavior of system can still be studied by numerical simulations.\par
In this new physical situation, the input parameters for the system are given by $A_0$ and $\omega$. The latter affects not only the periodicity of the local effect, but also its intensity. The limitation \eqref{eq:A0_k}, $A_0< k$, still apply.\par
\begin{figure}[H]
\captionsetup{skip=12pt,belowskip=15pt,font=small,labelfont=small,format=hang}
\centering
\begin{subfigure}{\textwidth}
\includegraphics[width=0.85\textwidth,keepaspectratio]{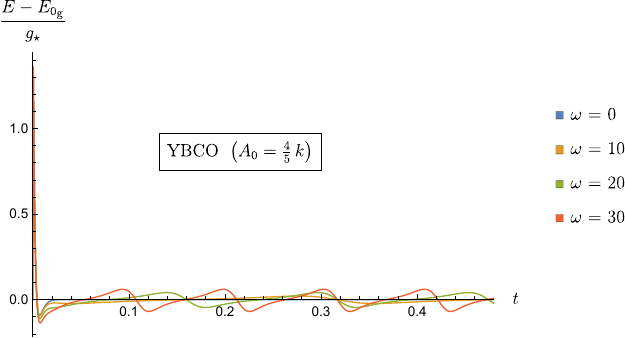}
\label{subfig:YBCO_Cos_A0small_wbig}
\end{subfigure}
\par\bigskip\bigskip
\begin{subfigure}{\textwidth}
\includegraphics[width=0.85\textwidth,keepaspectratio]{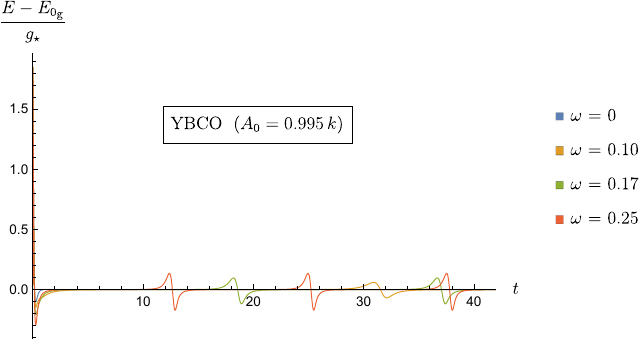}
\label{subfig:YBCO_Cos_A0big_wsmall}
\end{subfigure}
\par\bigskip\bigskip
\begin{subfigure}{\textwidth}
\includegraphics[width=0.85\textwidth,keepaspectratio]{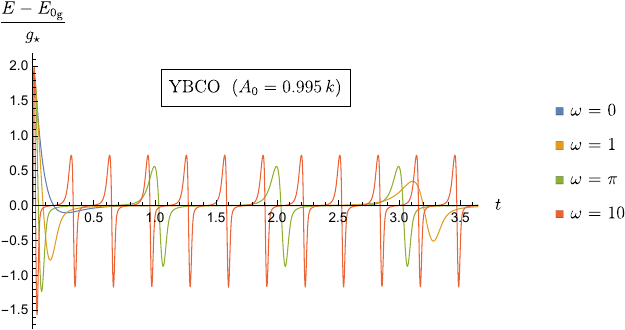}
\label{subfig:YBCO_Cos_A0big_wbig}
\end{subfigure}
\caption{Gravitational field alteration as a function of time (adimensional units) for an YBCO film subject to an oscillating external vector potential.}
\label{fig:YBCO_Cos}
\end{figure}
\begin{figure}[H]
\captionsetup{skip=15pt,belowskip=15pt,font=small,labelfont=small,format=hang}
\centering
\begin{subfigure}[t]{\textwidth}
\includegraphics[width=\textwidth,keepaspectratio]{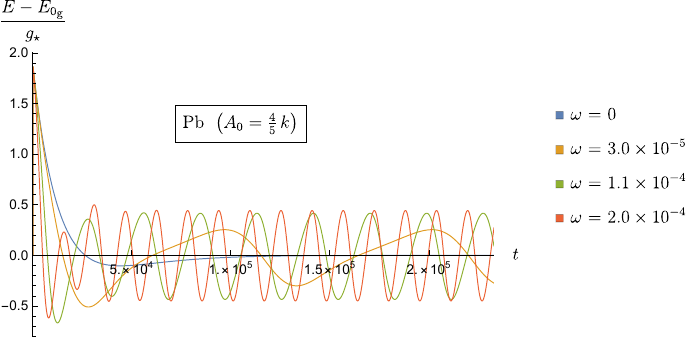}
\label{subfig:Pb_Cos_A0small_wsmall}
\end{subfigure}
\par\vspace{1.75cm}
\begin{subfigure}[b]{\textwidth}
\includegraphics[width=\textwidth,keepaspectratio]{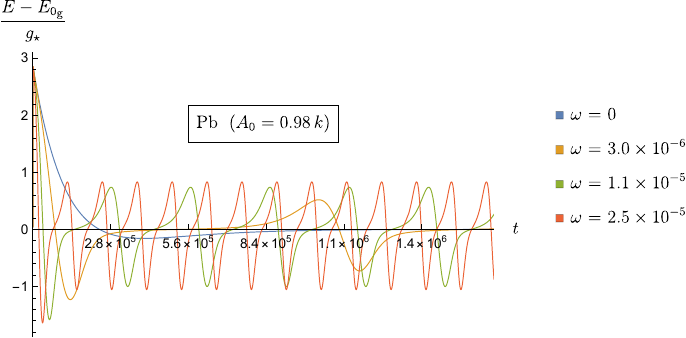}
\label{subfig:Pb_Cos_A0big_wsmall}
\end{subfigure}
\caption{Gravitational field alteration as a function of time (adimensional units) for a Pb film subject to an oscillating external vector potential.}
\label{fig:Pb_Cos}
\end{figure}

The results of the simulations are shown in Figures \ref{fig:YBCO_Cos} and \ref{fig:Pb_Cos}. As in the former case, the local gravitational field in the film is subject to positive and negative variations, that in this case are periodic in accordance to the form \eqref{eq:A0cos} of the vector potential.\par
From the same figures we can also see that the intensity of the local variation of the gravitational field increases as $k$ and $\omega$ increase. On the other hand, higher values of $\omega$ result in higher frequency rates of the oscillations: this in turn means that, from an experimental point of view, it would be more difficult to observe the phenomenon due to reduced time scales, since the average effect also tends to zero.\par
In view of the above discussion, the best physical setup for experimental detection involves an amplitude $A_0$ close to $k$ and a value of $\omega$ that is not too large. The latter choice is dictated by the discussed difficulties in time resolution for standard measurement systems.

\section{Concluding remarks}
\label{sec:concl}
In this work we have seen that suitable laser pulses can be used to maximize and amplify a local affection of the static Earth's gravitational field, in the simultaneous presence of a superconducting condensate. In particular, we have considered the vector potential associated to the laser pulse and its interaction with a superconducting thin film. This local interplay gives rise to a generalized electric-like field, featuring a gravitational component, and its explicit form can be obtained by studying the associated Ginzburg-Landau equations. We have then compared the latter perturbed field with the unperturbed value of the Earth's gravitational field, analysing the magnitude of the interaction and the time scales in which the phenomenon manifests itself.\par
A major challenge definetely consists in simultaneously optimizing the magnitude and duration of the effect, exploiting particular values of the input parameters, such as the intensity of the laser pulse and the physical characteristics of the superconductors. In this regard, most of the detection difficulties reside in reduced time scales for the occurrence of the local interplay. However, recent experimental setups should be able to probe effects taking place in time intervals well below nanoseconds \cite{mak2019attosecond}.

\bigskip
\bigskip
\section*{\normalsize Acknowledgments}
\vspace{-5pt}
We thank Fondazione CRT \,\includegraphics[height=\fontcharht\font`\B]{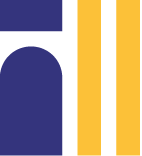}\:
that partially supported this work for A.\ Gallerati.
G.A.\ Ummarino acknowledges partial support from the MEPhI.

%

\bigskip
\begin{table}[H]
\centering
\begin{tabular}
{C{2cm} C{2.75cm} C{3.5cm}}
\toprule
\midrule
{ }  & \textbf{YBCO} & \textbf{Pb}
\\[4pt]
\midrule
\Tc & 89\,\Kelv & 7.2\,\Kelv
\\  \thinrule
T^{*} & 85\,\Kelv & 6.9\,\Kelv
\\  \thinrule
\xi(0) & 1.8\times10^{-9}\,\mt & 9.0\times10^{-8}\,\mt
\\  \thinrule
\xi(T^{*})& 8.5\times10^{-9}\,\mt  & 4.4\times10^{-7}\,\mt
\\  \thinrule
\lambda(0) & 1.7\times10^{-7}\,\mt & 4.0\times10^{-8}\,\mt
\\  \thinrule
\lambda(T^{*}) & 8.0\times10^{-7}\,\mt &  2.0\times10^{-7}\,\mt
\\  \thinrule
\Bctwo(0) & 61.0\,$T$ & 6.5\times10^{-2}\,$T$
\\  \thinrule
\Bctwo(T^{*}) & 2.5\times10^{-1}\,$T$ & 3.1\times10^{-3}\,$T$
\\  \thinrule
\mathcal{D} &  3.2\times10^{-4}\,\mt^{2}/$s$ & 1.0\,\mt^{2}/$s$
\\  \thinrule
k & 94.4 &  4.8\times10^{-1}
\\  \thinrule
\eta &  1.0\times10^{-3} &  6.6\times10^{3}
\\  \thinrule
\gstar(T^{*}) &  1.4\times10^{-10} &  1.2\times10^{-16}
\\  \thinrule
A_0^{{}^{\textsc{max}}}(T^{*}) &  2.8\times10^{-7}\,$V\,s/m$ & 8.8\times10^{-10}\,$V\,s/m$
\\  \thinrule
\tau(T^{*}) & 5.0\times10^{-10}\,$s$ & 3.2\times10^{-14}\,$s$
\\
\midrule
\toprule
\end{tabular}
\captionsetup{aboveskip=10pt}
\caption{Physical quantities related to YBCO and Pb.}
\label{tab:parameters}
\end{table}

\pagebreak


%
%

%
\appendix
\addtocontents{toc}{\protect\setcounter{tocdepth}{1}}
\addtocontents{toc}{\protect\addvspace{2.5pt}}%
\numberwithin{equation}{section}%

\section{Analytical solutions}
\label{app:sol}
Here we report the analytical form of the solutions \eqref{eq:fieldexpans} to the Ginzburg-Landau equations \eqref{eq:dpsi1}--\eqref{eq:dA2} for a vector potential of the form \eqref{eq:A0}:
\begin{equation}
\begin{split}
\psi_{1}(t)= &-1+\sqrt{\frac{k^{2}-A_{0}^{2}\:}{k^{2}-A_{0}^{2}\:\exp\left(-2\,k^{2}\,t-2\,A_{0}^{2}\:t\,\right)\;}}\;,
\\[1.275em]
\psi_{2}(t) =\,&\frac{\eta\,A_{0}}{\sqrt{k^{2}-A_{0}^{2}}}\:\frac{\:-k^{2}\exp\left(3\,k^{2}\,t\right)+\Big(\!-2\,A_{0}^{4}\,t+k^{2}\left(1+2\,A_{0}^{2}\,t\right)\Big)\,\exp\left(2\,A_0^2\,t\,+k^2\,t\right)\;}{\Big(k^{2}\,\exp\left(2\,k^{2}\,t\right)-A_{0}^{2}\,\exp\left(2\,A_{0}^{2}\,t\right)\Big)^{3/2}}\:,
\\[1.275em]
A_{1}(t) =\,&
A_{0}\:\theta(t)\,\exp\left(\!-\!\int_{0}^{t}\frac{\big(1+\psi_{1}(t)\big)^{2}}{\eta}\,dt\right)\:,
\\[1.5em]
A_{2}(t) =\:&\eta \,\exp\left(\!-\!\int_{0}^{t}\frac{\big(1+\psi_{1}(t)\big)^{2}}{\eta}\,dt\right)
-
\\[1ex]
&-2\int_{0}^{t} A_{1}\,\psi_{2}(t)\,\big(1+\psi_{1}(t)\big)\,
\exp\left(\int_{0}^{t}\frac{\big(1+\psi_{1}(t)\big)^{2}}{\eta}\,dt\right) dt\:.
\end{split}\end{equation}


\pagebreak
\hypersetup{linkcolor=blue}
\phantomsection 
\addtocontents{toc}{\protect\addvspace{3.5pt}}
\addcontentsline{toc}{section}{References} 
\bibliographystyle{mybibstyle}
\bibliography{bibliografia} 


\end{document}